\documentclass[12pt]{iopart}
\usepackage{graphicx} 
\usepackage{amssymb}
\usepackage[english]{babel}

\begin{document}
\title[Quantum Phase Slips in one-dimensional Josephson Junction Chains ] {Quantum Phase Slips in one-dimensional Josephson Junction Chains}

\author{Adem Erg\"ul\dag\ , Jack Lidmar\ddag\, Jan Johansson\S, Ya\u{g}{\i}z Azizo\u{g}lu\dag\, David Schaeffer\dag\ and David B. Haviland\dag}

\address{\dag\ Nanostructure Physics, Royal Institute of Technology, SE-106 91 Stockholm, Sweden}
\address{\ddag\ Theoretical Physics, Royal Institute of Technology, SE-106 91 Stockholm, Sweden } 
\address{\S Department of Natural Sciences, University of Agder, Kristiansand, Norway } 
 
\ead{adem@kth.se}
\begin{abstract}

We have studied quantum phase-slip (QPS) phenomena in long one-dimensional Josephson junction series arrays with tunable Josephson coupling. These chains were fabricated with as many as 2888 junctions, where one sample had a tunable weak link in the middle.  Measurements were made of the zero-bias resistance, $R_0$, as well as current-voltage characteristics (IVC).  The finite $R_0$ is explained by QPS and shows an exponential dependence on $\sqrt{E_J/E_C}$ with a distinct change in the exponent at  $R_0=R_Q=h/4e^2$.  When $R_0 > R_Q$ the IVC clearly shows a remnant of the Coulomb blockade, which evolves to a zero-current state with a sharp critical voltage as $E_J$ is tuned to a smaller value.  The zero-current state below the critical voltage is due to coherent QPS and we show that these are enhanced at the central weak link.   Above the critical voltage a negative differential resistance is observed which nearly restores the zero-current state.  

\end{abstract}
\submitto{\NJP}
\maketitle
%%%%%%%%%%%%%%%%%%%%%%%%%%%%%%%%%%%%%%%%%%%%%%%%%%%%%%%%%%%%%%%%%%%%%%%%%%%%%%%%%%%%%%%%%%%%%%%%%%%%%%%%%%%%%%%%%%%%%
%%%%%%%%%%%%%%%%%%%%%%%%%%%%%%%%%%%%%%%%%%%%%%%%%%%%%%%%%%%%%%%%%%%%%%%%%%%%%%%%%%%%%%%%%%%%%%%%%%%%%%%%%%%%%%%%%%%%%
%%%%%%%%%%%%%%%%%%%%%%%%%%%%%%%%%%%%%%%%%%%%%%%%%%%%%%%%%%%%%%%%%%%%%%%%%%%%%%%%%%%%%%%%%%%%%%%%%%%%%%%%%%%%%%%%%%%%%
%%%%%%%%%%%%%%%%%%%%%%%%%%%%%%%%%%%%%%%%%%%%%%%%%%%%%%%%%%%%%%%%%%%%%%%%%%%%%%%%%%%%%%%%%%%%%%%%%%%%%%%%%%%%%%%%%%%%%
\section{Introduction}

\indent An ideal superconductor has zero electrical resistance when it is cooled below a critical temperature $T_C$. However, a thin superconducting wire can exhibit non-zero resistance due to fluctuations of the superconducting order parameter which cause phase-slip events. The phase slip events can be thermally activated $TAPS$ (Thermally Activated Phase Slips)~\cite{McCumberPRB1970, LangerPR1967} or the result of quantum tunneling, $QPS$ (Quantum Phase Slip)~\cite{Giordano,Bezryadin,Arutyunov2008}. In this paper we examine QPS in long Josephson junction chains where the phase-slip rate and the spatial location of phase slips along the chain are experimentally well controlled.\\  

\indent QPS can be either incoherent or coherent.  The former are dissipative, where as the latter give rise to the Coulomb Blockade of Cooper pair tunneling  and Bloch Oscillations in a Josephson junction~\cite{AverinZh1985}, phenomena which are the electrodynamic dual to the DC and AC Josephson effects, respectively.  In thin films, this duality to superconductivity can lead to a 'superinsulator' state~\cite{Vikonur}. In nanowires there has been a long-standing interest in QPS and in the very existence of superconductivity in 1D systems~\cite{Giordano,LauNanowire, Bezryadin,Golubev}. Presently there is a growing interest in realizing coherent QPS in nanowires as a possible dual element to the Josephson junction~\cite{Nazarov, Astafiev,LehtinenPRL2012}. The Josephson junction series array is an artificial 1D system for the study of QPS, with great freedom of design.\\

\indent Series arrays have been previously studied as a model system for understanding QPS and in particular their role in quantum phase transitions~\cite{Bradley,Sondhi,FazioPR2001}. The series array can emulate an ideal superconducting nanowire when the array is long enough and uniform enough to hide its discrete nature, so that the probability per junction of QPS is relatively small and independent of position in the array. Classical simulations of phase slips in long arrays show that it is also necessary that the junction dynamics be overdamped in order for TAPS to occur uniformly along the array~\cite{ErgulPhaseStick}. Underdamped junction dynamics results in persistent phase slips at random nucleation sites, as opposed to random TAPS occurring with equal probability along the array. QPS have also been studied in short  arrays~\cite{HavilandJLTP,Pop2,Rastelli} and here we extend these studies to longer arrays. We present measurements of DC electrical transport in long arrays where we are able to  control the phase-slip rate uniformly along the array, and at one point in a the middle of the array. We show that the zero bias resistance $R_0$ is due to QPS.  When phase slips are localized to a central weak link, the rest of the array essentially acts as tunable environment for the study of localized QPS. Localized QPS in a circular array has recently been exploited in a promising new type of superconducting qubit~\cite{Manucharyan,Manucharyan2,KochPRX2013}.\\

\begin{figure}[t]
 \centering
     \includegraphics[width=0.43\textwidth]{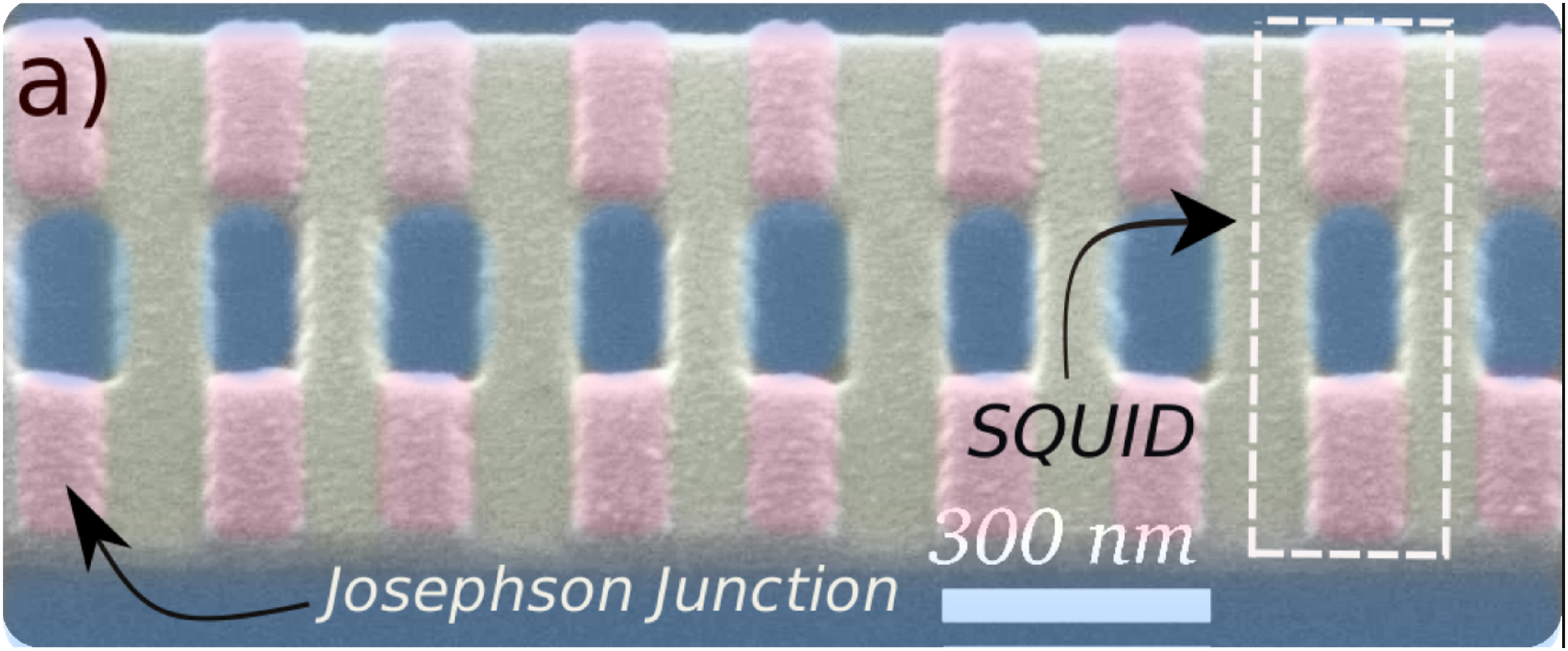}\\
     \includegraphics[width=0.44\textwidth]{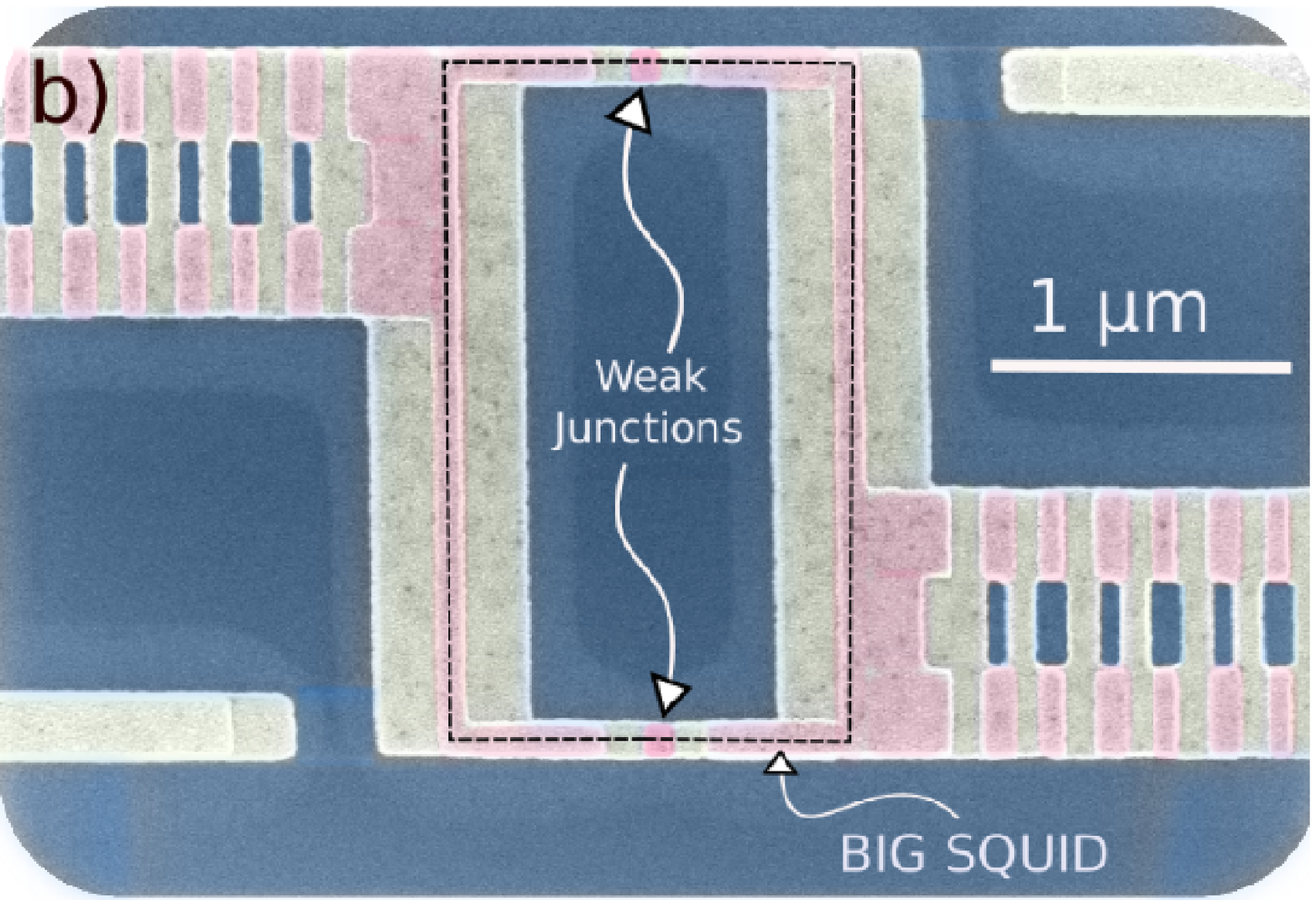}\\
\caption{a) Scanning Electron Microscope (SEM) image of a $uniform$ chain with the junctions sizes, $300~$nm $\times$ $100~$nm. b) SEM image of the $weak-link$ chain. Junctions of the big $SQUID$ have lower $I_C$ due to the smaller junction size, compared to the rest of the chain.}
 \label{fig:EBL}
\end{figure}

\begin{figure}[t]
\centering
\includegraphics[width=0.83\textwidth]{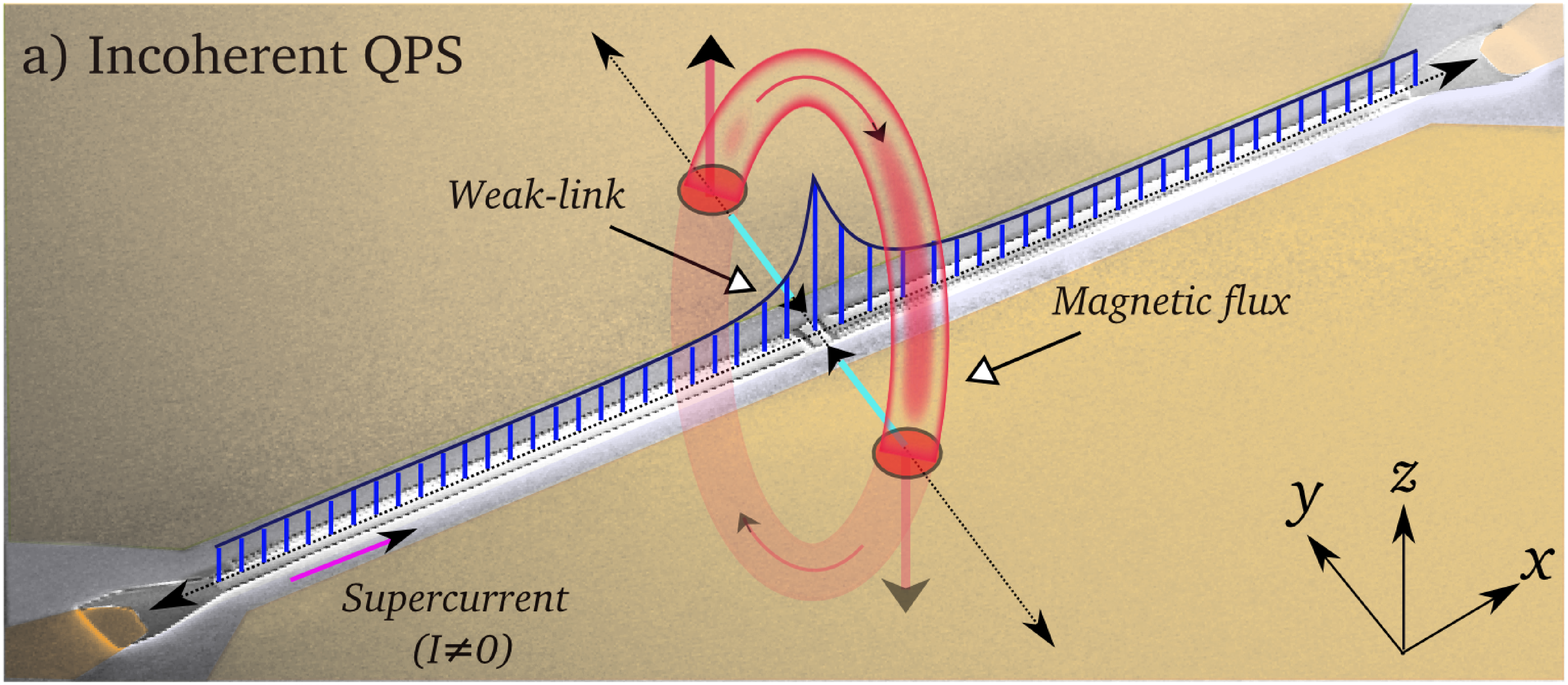}
\includegraphics[width=0.83\textwidth]{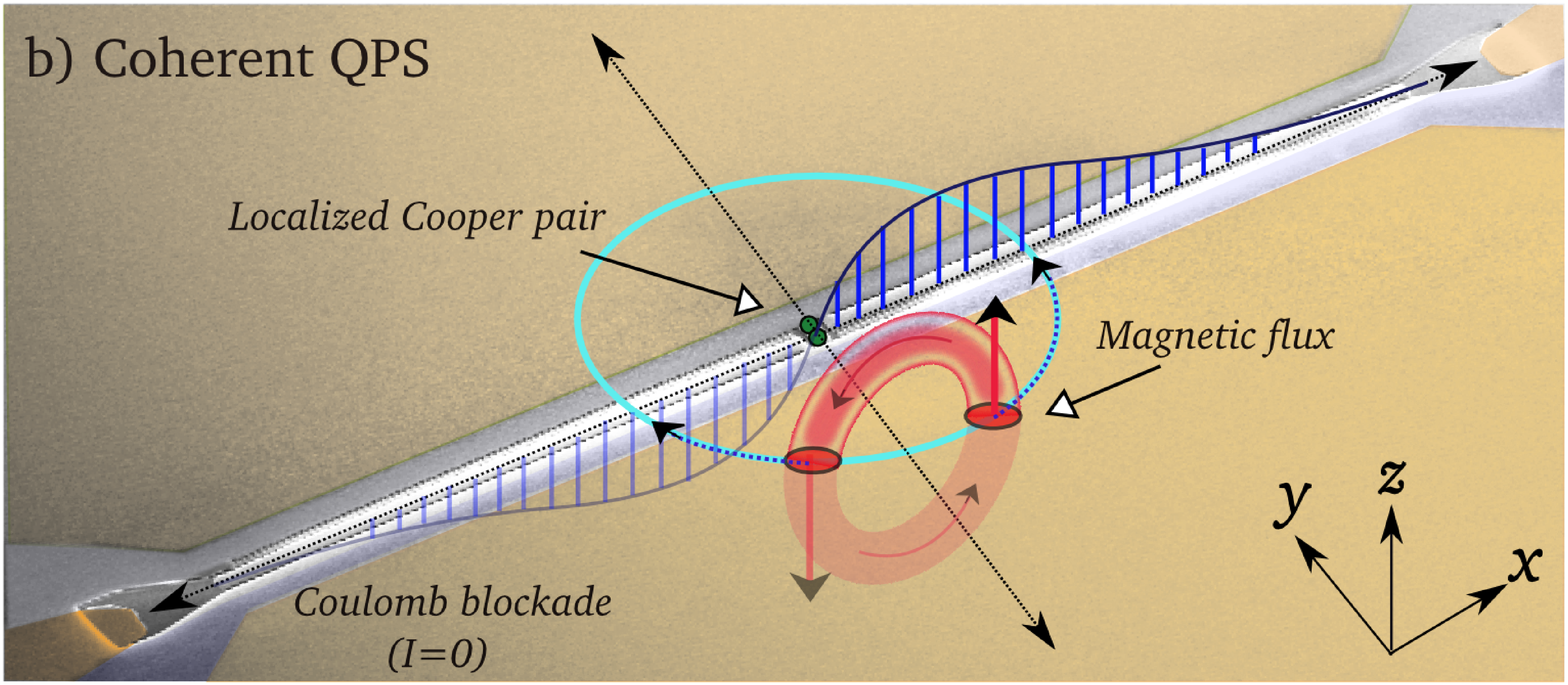}
\caption{Artist's conception of incoherent (a) and coherent (b) QPS events across a weak-link in a chain. The phase slip event is associated with the tunneling of a magnetic flux quantum (red tube) across the chain. The vertical blue bars represent the $x$ component of the electric field (potential gradient) due to the time average of many QPS.  Incoherent QPS with finite supercurrent give rise to dissipation, whereas coherent QPS describe a Coulomb blockade state, where the potential gradient is due to an excess, localized Cooper pair.}
\label{fig:Shape}
\end{figure}

\indent When the junctions in the array are formed as DC SQUIDs, the rate of QPS can be tuned with an external magnetic field~\cite{HavilandJLTP,Pop2,Rastelli}. The SQUID chain is described as a series array along the $x$ direction, where each junction is extended in the $y$ direction to form a loop with two parallel junctions (see figure~\ref{fig:EBL}). When an external magnetic field $B_z$ is applied in the $z$ direction the effective Josephson coupling between the series SQUIDs is tuned with the magnetic flux threading each loop, $E_J=E_{J0}|\cos(2\pi \Phi_{ext}/\Phi_0)|$. Here $\Phi_0=\frac{h}{2e}$ is the flux quantum and $E_{J0}=\frac{R_Q}{R_N}\frac{\Delta_0}{2}$ is the bare Josephson coupling determined by: the normal state resistance of each link in the chain $R_N$, the superconducting energy gap $\Delta_0$ and the quantum impedance  $R_Q=\frac{\Phi_0}{2e}=6.45\mathrm{k}\Omega$. The spatial dimensions transverse to the supercurrent flow are thus exploited to create a tunable 1D system. However, this tunable coupling requires that the loop inductance is small enough, $L_{\mathrm{loop}} E_{J0} \ll \left( \frac{\Phi_0}{2\pi} \right)^2$ such that the externally applied magnetic flux induces negligible circulating supercurrent in the loop, which is well satisfied in our chains.\\

\indent The charging energy, $E_C=e^2/(2C)$, is not considered as tunable and its value is fixed by the geometry of the islands in the chain. In this case the electrostatic energy associated with one uncompensated charge sitting in the middle of a long chain will depend on the entire capacitance matrix of the chain which, for simplicity, is usually assumed to have a symmetric tridiagonal form, $i.e.$ only nearest neighbor coupling, ~\cite{BakhvalovJETP1989,Likharev,0ConnellPRB1994}. The characteristic screening length of the field associated with this excess charge, or the charge soliton length $\Lambda = \sqrt{C/C_0}$, where $2C+C_0$ is the total capacitance of each island in the chain. In our experiments we have fabricated the chain as the center strip of a co-planar wave guide (see figure~\ref{fig:Shape}). We have estimated from geometry that $\sqrt{C/C_0} \cong 12$.\\

\begin{figure}[t]
\centering
\includegraphics[width=0.55\textwidth]{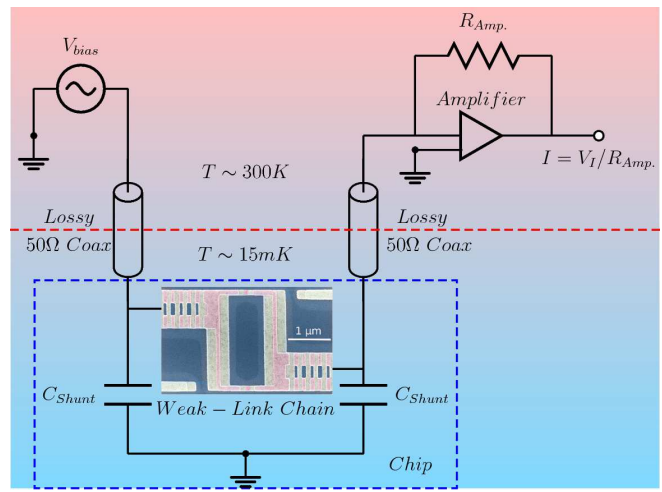}
\caption{Schematic diagram of the measurement circuit.}
\label{fig:MeasCircuit}
\end{figure}

\indent The superconducting state is characterized by a long range order of the phase of the complex order parameter. A supercurrent corresponds to a gradient of this phase and a phase-slip is an event in which the phase suddenly unwinds by $2\pi$ leading to a local decrease of the current and generation of a voltage pulse by the Josephson effects. We can also visualize a phase slip as a magnetic flux quantum crossing the chain, where the field associated with the flux, the direction of flux transport, and the direction of supercurrent flow, are all orthogonal (see figure~\ref{fig:Shape}). When there is a finite supercurrent, an incoherent QPS is associated with dissipation of electromagnetic energy stored in the fields surrounding the chain and it can be visualized as the disappearance of a circulating tube containing one quantum of magnetic flux (see figure~\ref{fig:Shape}a). In this process one flux quantum effectively crosses the chain, and the net effect of many random QPS is a time-average potential gradient at the crossing point, or component of the electric field $E_x$, parallel to the suppercurrent. This results in a Poynting vector with a component directed toward the chain.\\  

\indent Coherent QPS are associated with a zero-current state with no dissipation, and they describe a Coulomb blockade below a finite critical voltage. If an excess Cooper pair is placed in the center it will polarize the chain to form a charge soliton. The electrostatic potential gradient $E_x$ is positive on one side of the charge, and negative on the other (see figure~\ref{fig:Shape}b). If the potential gradient is interpreted as flux motion crossing the chain, it corresponds to a vacuum fluctuation where a tube of circulating flux is created on one-side of the chain, two opposing flux quanta traverse the chain on opposite sides of the charge, and are then annihilated. Thus a  virtual flux quantum effectively circulates once around the charge (Aharonov-Casher effect) beginning and ending with a state of zero flux. The sign of the potential gradient and the direction of circulation in the flux tube depend on the sign of the uncompensated charge quantum (i.e. charge soliton or anti-soliton).\\

%%%%%%%%%%%%%%%%%%%%%%%%%%%%%%%%%%%%%%%%%%%%%%%%%%%%%%%%%%%%%%%%%%%%%%%%%%%%%%%%%%%%%%%%%%%%%%%%%%%%%%%%%%%%%%%%%%%%%
%%%%%%%%%%%%%%%%%%%%%%%%%%%%%%%%%%%%%%%%%%%%%%%%%%%%%%%%%%%%%%%%%%%%%%%%%%%%%%%%%%%%%%%%%%%%%%%%%%%%%%%%%%%%%%%%%%%%%
%%%%%%%%%%%%%%%%%%%%%%%%%%%%%%%%%%%%%%%%%%%%%%%%%%%%%%%%%%%%%%%%%%%%%%%%%%%%%%%%%%%%%%%%%%%%%%%%%%%%%%%%%%%%%%%%%%%%%
%%%%%%%%%%%%%%%%%%%%%%%%%%%%%%%%%%%%%%%%%%%%%%%%%%%%%%%%%%%%%%%%%%%%%%%%%%%%%%%%%%%%%%%%%%%%%%%%%%%%%%%%%%%%%%%%%%%%%
\section{Experimental details}
\indent Long chains with as many as 4888 elements in series (9776 junctions - all working!) have been fabricated with electron beam lithography (EBL) by stitching several write fields together. Great care was taken in performing the write field alignment with a laser interferometer stage so that stitching errors did not effect the uniformity of junctions in the chain. The co-planar waveguide formed with these long chains has a microwave impedance which is much higher than the leads terminating its ends and therefore the chain is well voltage-biased at high frequencies relevant to phase dynamics. This impedance miss-match to the leads also means that internal junctions of the chain are rather immune to electromagnetic fluctuations in the leads.  However, to provide more filtering and a more perfect voltage bias, we fabricated a shunt capacitance to the ground planes (ca 180 pF) on chip, very close to each termination of the chain.  The first layer of the shunt capacitors are Al rectangles defined by optical lithography on Si/SiO$_2$ substrate. An insulating oxide (Al$_2$O$_3$) was then formed by heavy plasma oxidation and a second layer of photolithography and Au deposition formed the ground planes on-chip leads.\\

\begin{table}[t]
\centering
\footnotesize
\begin{tabular}{|c|c|c|c|c|c|c|c|c|c|c|c|c|}
\hline
$Parameters$  &$N$ &$A$&$R_{\mathrm{Tot}}$         &$R_{N}$        &$E_{J0}$ &$E_C$  & $E_{J0}/E_C$\\
\hline
$Units$ &\# &$\mathrm{\mu m^2}$			 		 &$\mathrm{M\Omega}$&$\mathrm{k\Omega}$  &$\mathrm{\mu eV}$ &$\mathrm{\mu eV}$&\# \\
\hline
$Sample~1$      &$2888$ &$0.05$  &$1.88$  &$0.65$  &$990$ &$35.6$  &$27.8$  \\
\hline
$Sample~2$           &$384$ &$0.06$  &$0.28$  &$0.72$  &$890$ &$29.6$  &$30$  \\ 
\hline              
$Sample~3$           &$2888$ &$0.05$  &$1.43$  &$0.5$  &$1302$ &$35.6$  &$36.6$  \\                   
\hline
$Sample~4$       &$2888$ &$0.02$  &$0.2$  &$0.0817$  &$7898$ &$88.8$  &$88.7$  \\
\hline
$Sample~5~(chain)$           &$384$ &$0.06$  &$0.86$  &$2.24$  &$288$  &$29.6$  &$9.7$  \\ 
\hline
$Sample~5~(weak-link)$           &$1$ &$0.02$  &$0.00672$  &$6.72$  &$96$  &$88.8$  &$1.08$  \\ 
\hline

\end{tabular}\\
\label{table1}
\caption{Parameters of the Josephson Junctions chains studied in this paper:  $N$ is the  number of series links in the chain; $A$ is the total area of two junctions in the loop which forms one link in the chain; $R_{\mathrm{Tot}}$ is the total normal state resistance of the chain; $R_{N}=R_{\mathrm{Tot}}/N$ is the normal state resistance of each link in the chain, assuming a uniform chain; $E_{J0}$ is the bare Josephson coupling energy of each link, determined from $R_{N}$ and the superconducting energy gap $\Delta_0$, $E_{J0}=(R_Q/R_{N})(\Delta_0/2)$;  $E_C$ is the charging energy of each link determined from the area $A$ and the specific capacitance $c_S=45~$fF/$\mu$m$^2$, $E_C=e^2/(2c_S A)$; $E_{J0}/E_C$ is the ratio between two characteristic energies at zero external magnetic field. The island stray capacitance is estimated to be $C_0 \approx 20$aF.}
\end{table}

\indent The lift-off mask for the Josephson junction chain was made by EBL using a two-layer resist system with over-development of the bottom layer to obtain large undercut and rather long ($\approx 1 \mathrm{\mu m}$) free-standing bridges.  The junctions were then fabricated by two-angle Al evaporation with {\it in situ} oxidation to form overlapping Al islands with an Al$_2$O$_3$ tunnel barrier~\cite{Dolan}. Numerous chains have been fabricated and measured, with a wide variety of parameters. In this paper we concentrate on describing measurements on 4 samples whose parameters are given in table 1. Samples 1-4 were uniform chains and a detail of the loop junctions can be seen in figure~\ref{fig:EBL}a. Sample 5 had a weak link fabricated at the center of the chain, which was realized by a loop with smaller junctions and larger loop area (figure~\ref{fig:EBL}b). The larger loop area allows for the coupling energy of the center link to be tuned with a period in magnetic flux that is different than the links in the chain.  Thus we can independently tune the coupling energies of both the single weak link, and the rest of the chain. The modulation of transport in the chain with magnetic field is periodic in $\Phi_0$ and we can use this fact to calibrate the modulation of $E_J$ directly as a function of the applied magnet current. Thus we can accurately determine the value of $E_J$ at any particular magnetic field for both the weak link and the chain links.\\ 

\indent After lift-off and initial testing the sample chip is wire-bonded to a printed circuit board with microwave connectors and mounted in a RF tight copper can.  Measurements were made in a dilution refrigerator with a base temperature of $\approx15$~mK. Figure~\ref{fig:MeasCircuit} shows a schematic diagram of the measurement circuit. The measurement leads from room temperature to the RF connectors on the copper can were ca 2.5 meters of continuous (no connectors) lossy micro coax cables with $50\Omega$ impedance.  These cables have poor thermal conductance (CuNi, 0.4 mm OD) and large microwave attenuation (61 dB/m @ 10 GHz) and thereby provide some thermalization of the microwave field fluctuations toward the base temperature of the cryostat.\\  

\begin{figure}[t]
\centering
\includegraphics[width=0.37\textwidth]{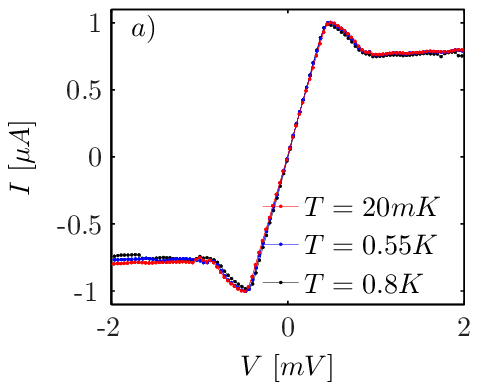}$~~$\includegraphics[width=0.38\textwidth]{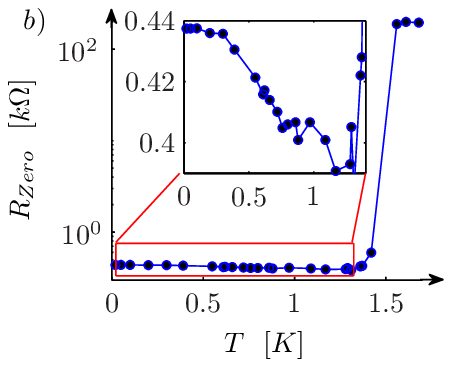}
\caption{ a) Various DC IV curves of Sample~4 between the $T=20~$mK to $T=0.8~$K at zero magnetic field. b) Zero bias resistance as a function of temperature and the inset shows the zoom to the $flat-tail$.}
\label{fig:RzeroIV}
\end{figure}

%%%%%%%%%%%%%%%%%%%%%%%%%%%%%%%%%%%%%%%%%%%%%%%%%%%%%%%%%%%%%%%%%%%%%%%%%%%%%%%%%%%%%%%%%%%%%%%%%%%%%%%%%%%%%%%%%%%%%
%%%%%%%%%%%%%%%%%%%%%%%%%%%%%%%%%%%%%%%%%%%%%%%%%%%%%%%%%%%%%%%%%%%%%%%%%%%%%%%%%%%%%%%%%%%%%%%%%%%%%%%%%%%%%%%%%%%%%
%%%%%%%%%%%%%%%%%%%%%%%%%%%%%%%%%%%%%%%%%%%%%%%%%%%%%%%%%%%%%%%%%%%%%%%%%%%%%%%%%%%%%%%%%%%%%%%%%%%%%%%%%%%%%%%%%%%%%
%%%%%%%%%%%%%%%%%%%%%%%%%%%%%%%%%%%%%%%%%%%%%%%%%%%%%%%%%%%%%%%%%%%%%%%%%%%%%%%%%%%%%%%%%%%%%%%%%%%%%%%%%%%%%%%%%%%%%
\section{Results and analysis}

\indent The current-voltage characteristics (IVC) of sample~4 at zero magnetic field and various temperatures are shown in figure~\ref{fig:RzeroIV}a. The IVC consist of a supercurrent-like branch with a finite slope $R_0$ at low bias voltages, followed by a peak and a constant current branch at high bias voltages.  The peak and constant current branch are reproduced by simulation of TAPS, where the constant current branch is due to a continuous and random slipping and sticking of the phase, occurring uniformly along the  chain~\cite{ErgulPhaseStick}. However, simulations show that the resistance of supercurrent-like branch, or $R_0$ due to TAPS, is much smaller than that measured in experiment. The measured temperature dependence of $R_0$ plotted in figure~\ref{fig:RzeroIV}b shows a nearly temperature independent 'flat tail' below the superconducting transition temperature of Al, with a slight {\em increase} in $R_0$ as temperature is decreased (figure~\ref{fig:RzeroIV}b inset). Resistance due to TAPS would have the opposite behavior of increasing $R_0$ with increasing temperature. The increase of $R_0$ at low temperatures in the chains can be qualitatively explained as being due to QPS within the context of a model for quantum phase transitions ~\cite{HavilandPC2001}, and a similar superconducting-insulating transition is also observed for thin superconducting wires~\cite{BezryadinJOP2008}.\\

\begin{figure}[t]
\centering
$~$\\
$~$\\
\includegraphics[width=0.6\textwidth]{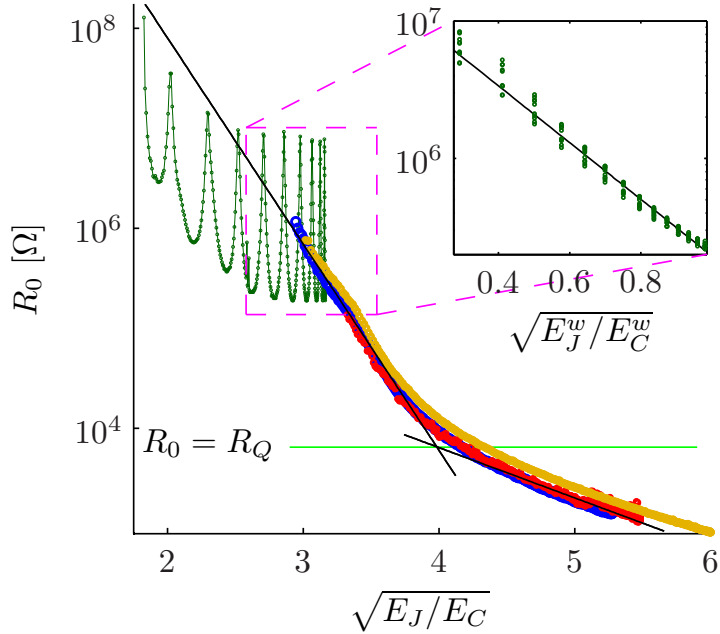}
\caption{Zero bias resistance plotted as a function of $\sqrt{E_J/E_C}$ for samples 1,2,3 and 5 (blue, red, orange and dark green points respectively). The black solid lines have slopes $\alpha_{\mathrm{low}}=1.16$ for $\sqrt{E_J/E_C} > 4$ and $\alpha_{\mathrm{high}}=4.75$ for $\sqrt{E_J/E_C} < 4$. Inset shows the $R_0$ modulation of the sample 5 as a function of  $\sqrt{E_J^w/E_C^w}$ of the weak-link. The slope of the black line in the inset is also $\alpha_{\mathrm{high}}=4.75$. }
\label{fig:RzeroTot}
\end{figure}

\indent The large value of $R_0$ and its weak temperature dependence tell us that the majority of the phase-slips are QPS~\cite{TempEst}. The tunneling amplitude $\Gamma$ of QPS depends on the ratio of the Josephson coupling energy to the charging energy, $E_J/E_C$~\cite{Matveev, Arutyunov2008}. In the case of a Josephson junction chain in the limit where $E_J \gg E_C$
 and $C \gg C_0$ one may estimate $\Gamma$ as~\cite{Matveev}
\begin{equation}
\Gamma = \sqrt{\frac 2\pi} N \hbar \Omega\sqrt{S}e^{-S},
\label{eq:1}
\end{equation}
\begin{equation}
S= \sqrt{8 E_J/E_C} ,
% S=\alpha \sqrt{E_J/E_C}
\label{eq:2}
\end{equation}
where $S$ is the action for an instanton connecting two neighbuoring minima and $\Omega=\sqrt{8E_J E_C}/\hbar$ is the plasma frequency of the Josephson Junction. Interestingly the same expression for $\Gamma$ holds for a single junction if one omits the $N$ in the prefactor. For finite $C_0$, and long chains with $N \gg \Lambda$, one expects an additional contribution $\sim \pi \sqrt{E_J/8E_C \Lambda^2} \ln (N/\pi \Lambda)$ to $S$, which is negligible when $\Lambda \gg 1$~\cite{Rastelli}.\\

\indent To explain our measurements of the zero bias resistance $R_0$ we make the reasonable assumption that  $\log R_0 \propto S$ so that in the limit $E_J/E_C \gg 1$, we would have  
\begin{equation}
   R_0 \propto e^{-\alpha \sqrt{E_J/E_C}},
   \label{eq:3}
\end{equation}
where we treat $\alpha$ as a fitting parameter. Figure~\ref{fig:RzeroTot} displays $\log R_0$ vs. $\sqrt{E_J/E_C}$  for samples 1-3 which are uniform chains, and for sample 5 which contains the weak link in the middle of the chain. The uniform chains do indeed follow the exponential dependence of equation~(\ref{eq:3}) in the large $E_J$ limit. Surprisingly, the same exponential relation holds in the lower $E_J$ limit, but with a different $\alpha$ value. We observe a clear transition between different exponential factors $\alpha_{\mathrm{low}}=1.16$ and $\alpha_{\mathrm{high}}=4.75$, when the zero bias resistance approximately equals the resistance quantum ($R_0 \approx R_Q$). This happens around $\sqrt{E_J/E_C} \approx 4$. Note also that although these uniform chains have very different number of junctions $N$, they have essentially identical dependence of $R_0$ on $\sqrt{E_J/E_C}$. Apparently these chains are not influenced by chain lengths which is contrary to earlier $R_0$ measurements on shorter chains~\cite{HavilandPC2001}. Theoretically one would expect the length enter into the prefactor in equation~(\ref{eq:3}).\\

\begin{figure}[t]
\centering
\includegraphics[width=0.6\textwidth]{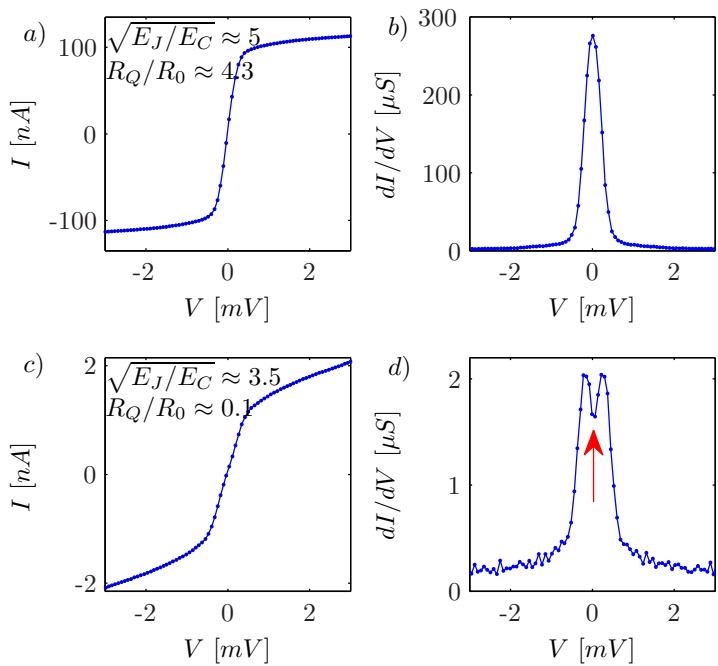}
\caption{ DC IV curve (a) and differential conductance (b) of sample~2 as a function of bias voltage at $\sqrt{E_J/E_C}\approx 5$ and  $R_Q/R_0 \approx 4.3$. DC IV curve (c) and differential conductance (d) of the same sample at $\sqrt{E_J/E_C}\approx 3.5$ and  $R_Q/R_0 \approx 0.1$. Differential conductance curve shows the suppression of the zero bias conductance and the remnant of the Coulomb blockade state (red arrow).}
\label{fig:DCIVsamp2}
\end{figure}

\indent Also shown in figure~\ref{fig:RzeroTot} is sample 5 with the tunable weak link. The modulation of $R_0$ seen in this sample is due to modulation of $E_J$ of the weak link junction, whereas the change in the average $R_0$ is due to the chain. Since the modulation of $E_J$ is much faster for the weak-link junction than for the chain junctions, the modulation of $R_0$ becomes quasi-periodic when plotted against $\sqrt{E_J/E_C}$ of the chain. In the region where $\sqrt{E_J/E_C}<2.5$ the maximum points of the modulated $R_0$ fall close to the line with the same value of $\alpha_{\mathrm{high}} = 4.75$, indicating that at these maxima, QPS in the chain is dominating the transport. For $\sqrt{E_J/E_C}>2.5$ the zero bias resistance is mostly influenced by the dynamics of the weak link. Plotting log $R_0$ vs. $\sqrt{E_J^w/E_C^w}$ of the weak-link junction in the inset of figure~\ref{fig:RzeroTot} we see an exponential behavior according to equation~(\ref{eq:3}) with the same slope  $\alpha_{\mathrm{high}}=4.75$.\\

\indent The transition around $R_Q$ with its associated change in the exponent $\alpha$ is also accompanied by a qualitative change in the nonlinear character of the IVC.  Figure~\ref{fig:DCIVsamp2} shows the IVC and the differential conductance of sample 2 measured at either side of the transition. When $R_0 < R_Q$, the differential conductance displays a single peak at zero bias (figure~\ref{fig:DCIVsamp2}b). However, when $\sqrt{E_J/E_C}$ is tuned so that $R_0>R_Q$, we see a dip in the differential conductance at zero bias (figure~\ref{fig:DCIVsamp2}d). This dip is a remnant of the Coulomb blockade which is heavily smeared by quantum fluctuations of the quasicharge (incoherent Cooper pair tunneling)~\cite{IngoldNazarov}. As the magnetic field is increased and $E_J$ is suppressed further, this dip develops into a true zero current state with an $E_J$ dependent critical voltage.\\ 

\begin{figure}[t]
\begin{center}
\includegraphics[width=0.8\textwidth]{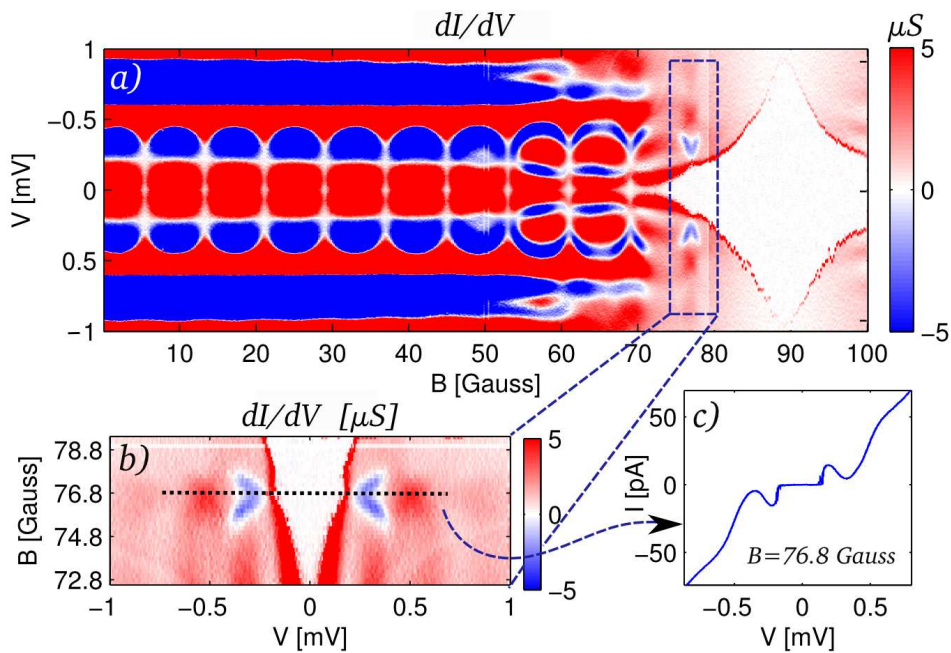}
\caption{a) Color map of the differential conductance of sample 5 as a function of magnetic field and bias voltage. The periodic pattern with period 7.5 G corresponds to one flux quantum in the large loop of the weak-link junction. b) Zoom to the region where the chain impedance increase such that the current-voltage characteristics is determined by both the chain and the weak-link. It is also possible to see the enhancement of the coulomb gap due to the weak-link at B$\approx 76.8~$G. c) DC IV curve of the sample at B=$76.8~$G.}
\label{fig:RzeroB}
\end{center}
\end{figure}

\indent The samples 1-3 showed this transition in the IVCs as the zero-bias resistance crossed $R_Q$. However, the weak-link sample (sample 5) showed a zero-bias dip in the differential conductance at \textit{all} magnetic fields due to the small $\sqrt{E_J^w/E_C^w}$ value of the weak-link junction. Sample 5 has two different loop sizes and its differential conductance therefore shows a rich dependence on magnetic field and bias voltage, shown as a color map in figure~\ref{fig:RzeroB}a. The periodic pattern with period $7.5~$G corresponds to one flux quantum in the larger loop of the central weak link (see figure~\ref{fig:EBL}b). Below $50~$G a small zero-bias dip in differential conductance is most clearly visible at magnetic fields were the weak link coupling is suppressed to a minimum. At magnetic fields above $50~$G, $E_J$ of the chain junctions begins to be suppressed, causing an increase in the chain impedance. We see that the small dip in differential conductance at zero bias becomes extended to higher voltage as the Coulomb blockade in the weak-link is enhanced. Above $70~$G the Coulomb blockade in the chain starts to take over with a rapidly increasing critical voltage which reaches a maximum at $90~$G, where $E_J$ of the chain junctions is at a minimum.\\ 

\indent A very interesting feature occurs at $76.8~$G, where the weak link coupling is at a minimum and the chain has a well-developed Coulomb blockade (see figure~\ref{fig:RzeroB}b). The critical voltage shows a small increase at this magnetic field, signaling that the weak link is contributing by a noticeable amount to the total critical voltage. The IVC at this magnetic field is shown in figure~\ref{fig:RzeroB}c and it displays a well-developed zero-current state and distinct critical voltage. When the chain is biased above the critical voltage there is a sharp onset of Cooper pair tunneling, followed by a region with negative differential resistance (NDR) reaching a nearly zero-current state. In the zero current state the QPS act coherently to create an insulating Coulomb blockade for Cooper pair tunneling. The onset of current is associated with Bloch oscillations which are somewhat enhanced at the weak-link. The Bloch oscillations in the center of the chain are happening in the presence of the complicated nonlinear electromagnetic environment of the rest of the chain. Noteworthy is the region with NDR, as it suggests that these Bloch oscillations are exciting a low-dissipation mode of the environment. 

%%%%%%%%%%%%%%%%%%%%%%%%%%%%%%%%%%%%%%%%%%%%%%%%%%%%%%%%%%%%%%%%%%%%%%%%%%%%%%%%%%%%%%%%%%%%%%%%%%%%%%%%%%%%%%%%%%%%
%%%%%%%%%%%%%%%%%%%%%%%%%%%%%%%%%%%%%%%%%%%%%%%%%%%%%%%%%%%%%%%%%%%%%%%%%%%%%%%%%%%%%%%%%%%%%%%%%%%%%%%%%%%%%%%%%%%%

\section{Conclusion}

\indent We have found evidence for incoherent QPS in long Josephson junction chains. The zero bias resistance is well described by simple exponential behavior, $R_{0} \propto \exp(-\alpha \sqrt{E_J/E_C})$. This dependence is consistent with theoretical estimates based on QPS in the limit of large $E_J/E_C$ and vanishing ground capacitors. Interestingly, a crossover to another regime ($E_J \lesssim 16E_C$) with the same type of exponential behavior, but a higher value of $\alpha$, occurs around the resistance quantum $R_Q=h/4e^2$.
In the IVC the crossover is accompanied by the appearance of a zero bias anomaly which is a remnant of the Coulomb blockade of Cooper pair tunnelling.  This qualitiative change at $R_0=R_Q$ represents a crossover to a regime where QPS are strongly interacting and start to develop coherence. A theory for this regime is currently lacking. Although the chains are very uniform and QPS could happen anywhere, we observe no strong length dependence neither in the exponent $\alpha$ nor in the prefactor of the exponential. Previous measurements on shorter chains showed  very clear finite size effects~\cite{HavilandPC2001}. The reasons for this unexpected absence of length dependence is not understood.\\

\indent As $E_J$ is further suppressed the zero-bias anomaly in the IVC eventually develops into a zero-current state with a critical voltage that depends on $E_J$. We find that when this critical voltage is enhanced by an increase of coherent QPS at the weak link, there appears a very interesting NDR in the IVC above the critical voltage.  We speculate that this NDR is due to Bloch oscillations in the weak link, which are exciting a low-dissipation mode of the chain.  Understanding the nature of this mode could provide insight as to how one might structure coherent QPS along the chain and provide a clue as to how to design a chain where Bloch oscillations can be synchronized to an external electromagnetic signal.\\

%%%%%%%%%%%%%%%%%%%%%%%%%%%%%%%%%%%%%%%%%%%%%%%%%%%%%%%%%%%%%%%%%%%%%%%%%%%%%%%%%%%%%%%%%%%%%%%%%%%%%%%%%%%%%%%%%%%%
%%%%%%%%%%%%%%%%%%%%%%%%%%%%%%%%%%%%     BIBLIOGRAPHY     %%%%%%%%%%%%%%%%%%%%%%%%%%%%%%%%%%%%%%%%%%%%%%%%%%%%% 
 %%%%%%%%%%%%%%%%%%%%%%%%%%%%%%%%%%%%%%%%%%%%%%%%%%%%%%%%%%
%\def\newblock{\hskip .11em plus .33em minus .07em}


\begin{thebibliography}{<31>}
\bibitem{TempEst}
In the case of {TAPS} we would expected to see a different kind of exponential
  dependence, {$R_{Zero} \propto \exp(-2 E_J/(k_BT))$}. {B}y fitting the
  {$R_{0}$} vs {$E_J$} data of sample 2 we get {$T\approx 6.7K \gg T_C$}.
\end{thebibliography}
\end{document}